\documentclass[prl,showpacs,twocolumn,amsmath,amssymb,superscriptaddress]{revtex4}%
\usepackage{graphicx}% Include figure files
\usepackage{dcolumn}% Align table columns on decimal point
\usepackage{colortbl}
\usepackage{bm}% bold math\newcommand{\kvec}{{\bf k}}
\newcommand{\rvec}{{\bf r}}

\newcommand{\Qvec}{{\bf Q}}

\newcommand{\beq}{\begin{equation}}
\newcommand{\eeq}{\end{equation}}
\newcommand{\beqa}{\begin{eqnarray}}
\newcommand{\eeqa}{\end{eqnarray}}

\begin{document}
\title{Hidden Ferronematic Order in Underdoped Cuprates}
\author{G. Seibold}
\affiliation{Institut f\"ur Physik, BTU Cottbus, PBox 101344, 03013 Cottbus, 
Germany}
\author{M. Capati}
\author{C. Di Castro}
\author{M. Grilli}
\author{J. Lorenzana}
\affiliation{ISC-CNR, CNISM and
Dipartimento di Fisica, Universit\`a di Roma ``La Sapienza'', P.le Aldo
Moro 5, I-00185 Roma, Italy}
\date{\today}% It is always \today, today,
\begin{abstract}
We study a model for low doped  cuprates where holes 
 aggregate into oriented stripe segments which have a vortex and an
 antivortex fixed to the extremes. 
We argue that due to the interaction between segments 
a state with macroscopic polarization is stabilized,  
which we call a ferronematic. This state can be characterized as a 
charge nematic which, due to the net
polarization, breaks inversion symmetry and also exhibits  an 
incommensurate spin modulation. Our calculation can reproduce the  
doping dependent spin structure factor of lanthanum cuprates in
excellent agreement  with experiment and allows to rationalize
experiments in which the incommensurability has an order
parameter-like temperature dependence.  
\end{abstract}


\maketitle
The question whether there is a broken symmetry hidden order in 
the pseudogap phase of cuprate high-temperature superconductors 
is still a matter of debate. In this regard, a popular proposal is unidirectional
charge\cite{eme93,cas95} or spin and charge order\cite{zaa89,mac89}
known as stripes.  

Bulk evidence for long-range charge stripe order 
has only been seen in few compounds.~\cite{tra95,abb05,fink09}
La$_{2-x}$Sr$_x$CuO$_4$ (LSCO) shows low energy
incommensurate spin scattering, usually associated with some kind of
stripe correlation,  which undergoes a rotation towards 
the diagonal direction below
hole concentrations $x\lesssim 0.05$, where it even becomes
elastic.~\cite{waki99,waki00,mats00} The 
incommensurability $\delta$ (proportional to the inverse of the 
spin modulation) evolves linearly $\delta \sim x$ 
from $x=0.02$ up to the metallic phase where it saturates above $x
\gtrsim 1/8$\cite{yam98} as expected for stripes.~\cite{lor02}

Quasistatic incommensurate spin scattering along the Cu-O bond direction
has also been found in detwinned YBCO.~\cite{hinkov,hau10} 
In this case, the absence of signatures of long-range charge order and the 
evidence of rotational symmetry breaking \cite{hinkov,hau10,daou10} points 
towards a nematic order. It has been proposed that  this 
order arises either from incipient unidirectional  fluctuating stripes 
\cite{vojta} or  from an independent d-wave type nematic actor which preserves
translational symmetry.~\cite{sun10} 
Other proposals for a broken symmetry state include orbital currents
which break time reversal symmetry\cite{var99}, spirals\cite{sus05} 
and a d-density wave.~\cite{ben00,cha00}

\begin{figure}[tb]
\includegraphics[width=8.5cm,clip=true]{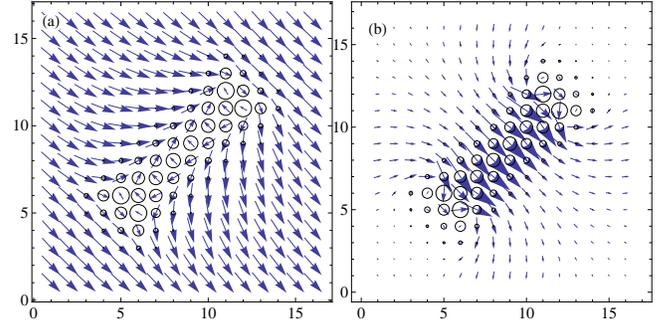}
\caption{(color online). a) Stripe segment for $8$ holes (corresponding to 
4 VA pairs) obtained by minimizing the GA 
energy of  on a $16\times 16$ lattice in the one band Hubbard
models with $t'/t=-0.2$.  The radius of the circles are
proportional to the added hole density while the arrows are the
staggered magnetization.  b) Spin currents defined from the
conservation of the $z$ component of the
magnetization\cite{sei982}. In the continuum limit the spin current
is proportional to the gradient of the phase of the staggered magnetization.   }
\label{fig4}
\end{figure}

In this paper we show that neutron scattering experiments in  strongly
underdoped cuprates can be understood in terms of 
a phase which breaks  rotational {\em and} 
inversion symmetry. It is formed by oriented stripe segments which do not
need to have positional order thus we call it ferronematic. The segments are
oriented because they sustain a vortex and an antivortex of the
antiferromagnetic (AF) order in the extremes (Fig.~\ref{fig4}). Although formally the phase
has zero ordering wave vector in the charge sector, we show that it induces 
magnetic incommensurate peaks in excellent agreement with experiment
(Fig.~\ref{fig6}).  Remarkably, the order parameter is proportional to
the incommensurability as suggested by experiment.~\cite{hinkov,hau10} 

We focus on non-superconducting underdoped LSCO
which can be grown in a structure with only two twin domains with
different population. Therefore similarly to the case of YBCO the 
one-dimensionality of the low energy (diagonal) spin response can be 
clearly resolved.~\cite{waki00,mats00}

Previous variational computations
\cite{verges,sei982,berciu,koizumi}  
for the (low) doping induced spin structure in cuprates suggest the formation 
of magnetic vortex-antivortex (VA) pairs the relevance of which has been
discussed e.g. in context of the destruction of long-range  N\'{e}el 
order.~\cite{timm} In order to study these textures we  
have performed variational calculations based on the Gutzwiller
approximation (GA) of the extended one-band Hubbard model.
The ratio between on-site repulsion
$U$ and nearest-neighbor hopping $t$ is set to $U/t=8$ as suggested by
previous studies.~\cite{lorcol,sei05} The ratio between next-nearest
neigbor hopping $t'$ and $t$ is taken as a material parameter.~\cite{pav01}

Within this framework the energetically most 
stable solution for two holes is a VA
pair.~\cite{sei982} Our GA computations reveal that the interaction
between the pairs is dominated by an anisotropic   
short-range core-core contribution which originates from the
distribution of the localized holes.  This results in a head-to-tail
aggregation of VA pairs which tend to form chains in such a way that
only the vortex and the antivortex on the extremes contribute to the
long range distortion.   
 Fig.~\ref{fig4}(a) shows the spin and 
charge structure for  8 holes corresponding to 
4 VA pairs. Examination of the spin current (b)
allows to visualize the VA pair nucleated at the extremes and the fact
that the texture breaks inversion symmetry.  
Notice that the segments tend to form an antiphase domain wall of the
AF order although the transition from finite segments
to infinite stripes is non trivial and will be discussed elsewhere.~\cite{goe12}

\begin{figure}[tb]
\includegraphics[width=8cm,clip=true]{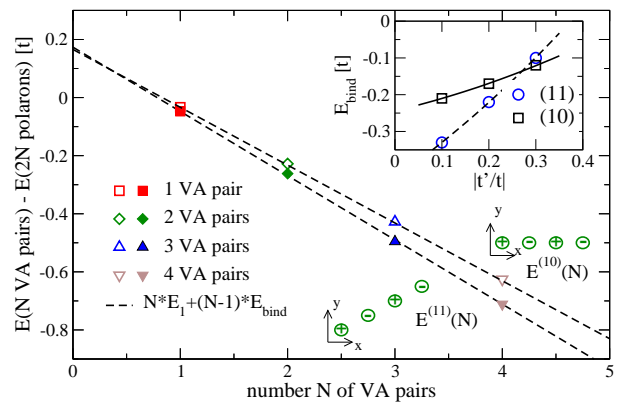}
\caption{(color online). Energy of a chain of $2N$ holes corresponding to  
$N$ V-A pairs  aligned along
the vertical (open symbols) and diagonal direction (full symbols) 
as compared to the energy of $2N$ polarons. The slope of the
dashed lines corresponds to the binding energy between pairs.
Parameters: $U/t=8$, $t'/t=-0.2$. The upper right inset shows the binding
energy as a function of $t'/t$ for vertical (squares) and diagonal (circles)
directions. Lines are guide to eyes. Computations where done in
systems with up to $20 \times 20$ sites. 
}
\label{fig3}
\end{figure}

Fig.~\ref{fig3} shows that segments formed by $2N$ holes have
systematically lower energy than $2N$ spin polarons with the preferred
orientation depending on parameters as discussed below. Seen the segment
as formed by $N$ VA pairs we obtain the binding energy $E_{bind}$
between the pairs from $E(N)=N*E_1+(N-1)E_{bind}$.
Parameters appropriate for lanthanum cuprates 
($t'/t \sim -0.15 \dots\! -0.2$) \cite{pav01},
yield a slight preference for diagonally oriented segments. In addition we have checked 
that in the low temperature orthorhombic (LTO) phase an anisotropy of 
$t'/t$ along orthorhombic a- and
b-axis favors the orientation of the VA-pairs along the a-axis.

Until now we have neglected the long-range part
of the Coulomb interaction  which limits the infinite
aggregation of VA pairs. In the spirit of Ref.~\cite{lor01}
we may estimate its effect by considering
the segments (each hosting $N_c$ charges) embedded in a homogeneously 
charged background. For large $N_c$ the Coulomb energy per charge 
increases logarithmically with the number of holes. At a fixed doping
concentration $x$ the energy per planar Cu reads,  
\begin{equation}\label{eq:coul}
\frac{E(N_c)}{L^2} = x \left[ E_c \ln(N_c) +
  \gamma + \frac1{N_c} |E_{bind}|\right]
\end{equation}
with $E_c=e^2/(\epsilon_0 a_{o})$ a charging energy expressed in
terms of the orthorhombic lattice constant $a_{o}$, the static dielectric
constant $\epsilon_0$, and $\gamma=(E_1+E_{bind})/2$. The last term comprises the fact that
shorter segments have less binding energy.  
The energy is minimized by $N_c=|E_{bind}|/E_c$ which leads to short
segments of only few lattice constants at infinitesimal doping. 
Considering a Wigner crystal
of segments and screening we find that the length of the segments tends
to grow rapidly with doping and may even diverge at 
a critical doping.\cite{goe12}

It is useful to remember that a XY
vortex can be mapped to a 2D Coulomb charge\cite{min87} or alternatively 
to current wires perpendicular to the 2D plane. The sign of the
charges or the direction of
the current is then determined by the vorticity.  
Thus the VA pairs map into 2D electric
dipoles in the Coulomb analogy or 2D magnetic dipoles perpendicular to
the former in the current analogy. In the latter case the magnetic
field maps into the spin current which will be used below. 

We assume that quenched disorder will yield a state in which segments
have no positional order. We have checked numerically \cite{goe12} 
that the combination of long-range dipolar interaction plus short
range interaction favors a ferronematic alignment of the dipoles.

Information on the spin response of a large aggregate of segments 
 from the GA is hampered by the finite size of the clusters. Since the
 textures are planar and we are interested in the large scale behavior we
consider  a classical AF $XY$-model with nearest neighbor
interaction $J$.
The segments are modeled as a chain of vacancies which
alternately correspond to the center of a vortex and antivortex. 
The spin structure is then determined from the minimization of the 
classical energy. In order to prevent annihilation of VA pairs we introduce 
a frustrating AF second next nearest neighbor coupling $J'$ across the
segments. The value of $J'$ can be obtained by 
comparing the phase change across a single segment with the GA calculation
(Fig. \ref{fig4}) which yields $J' \approx J$.

As an additional parameter we have to fix the filling factor $\nu$
which is the number of charges per segment length. 
Whereas the site-centered VA chain
shown in Fig. \ref{fig4} has $\nu \approx 1$ there exists a similar 
plaquette centered structure with comparable (though slightly higher)
energy, which has $\nu \approx 0.65$. The filling factor of these structures
is thus similar to those of (infinitely extended) diagonal 
site- and bond-centered stripes. \cite{goe09} The following calculations
assume $\nu=0.7$. Note, however, that experimental data
imply an increase towards $\nu=1$ close to the AF
boundary.~\cite{mats00}

\begin{figure}[bt]
\includegraphics[width=8.5cm,clip=true]{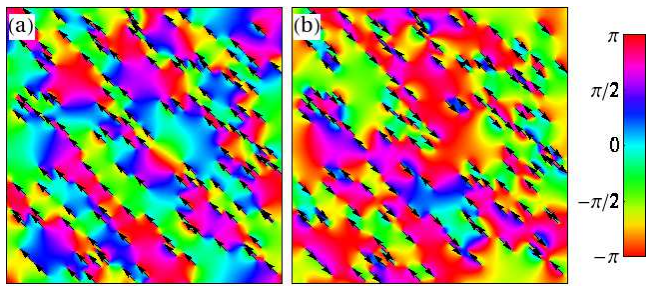}
\caption{(color online). 
Spin phase distribution  for 
(a) macroscopically and 
(b) randomly polarized distribution of stripe dipoles in a
160$\times$160 sites system.
Segments have a length of $8$ sites on the diagonal 
(4 VA pairs) and are represented by arrows indicating the length and 
polarization. With a filling factor
 $\nu=0.7$ the total number
of segment sites corresponds to a doping of $x=0.03$.}
\label{fig5}
\end{figure}

\begin{figure}[tb]
\includegraphics[width=8cm,clip=true]{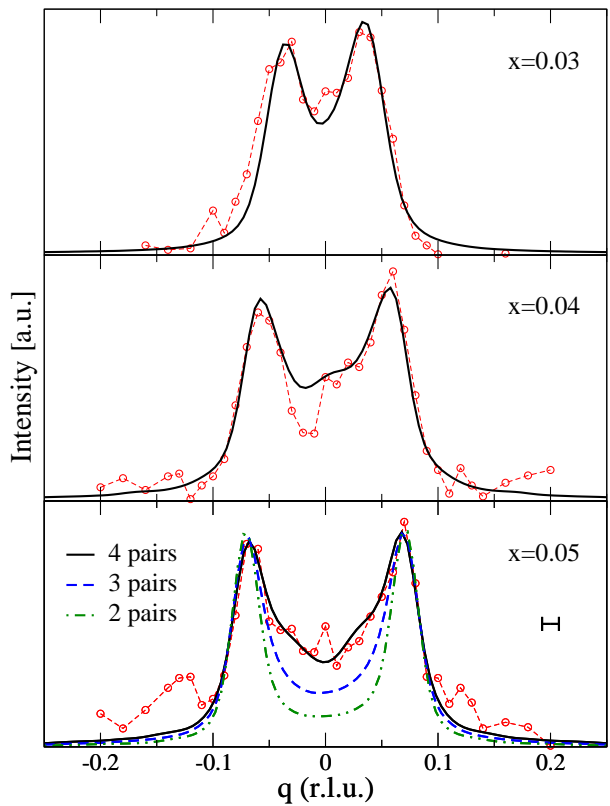}
\caption{(color online). Fits of the spin structure factor (LSCO) at different dopings 
for $N_c=8$ segments as explained in the text. 
For $x=0.05$ we also show spectra for $N_c=4,6$ segments for comparison. 
Computations have been done on lattices with up to $160 \times 160$ sites 
and we average over $20-30$ segment configurations where the 
experimental resolution (horizontal bar in the lower panel, cf.
Fig. $4$ of Ref. \onlinecite{waki00}) has been taken
into account by convoluting with a gaussian. 
Data by courtesy of S. Wakimoto.}
\label{fig6}
\end{figure}

Fig.~\ref{fig5}(a) reports the spin phase distribution for a particular
distribution of stripe dipoles with segment length of $8$ sites at $x=0.03$, 
all polarized along the
$[-1,1]$-direction. 
One observes a monotonous increase on the phase of the staggered
magnetization along the $[1,1]$-direction, although the distribution
of stripe segments is completely random. 
In Fig.~\ref{fig5}(b),  for the same distribution of segments, 
the associated dipole orientation
is now completely random. In this case the system disaggregates into
large areas with similar phase. 

The numerical results can be understood by considering the long-range 
distortion produced by the segments. Using the magnetic analogy
mentioned above and 
standard arguments from the
theory of macroscopic dielectrics and magnetized systems\cite{jac98}
it is easy to 
see that the system develops a macroscopic ``magnetic field''
equivalent to an average net spin current
perpendicular to the segments orientation and proportional 
to the average polarization. 

Analogously we can  obtain this result by considering 
the influence of a collection of equally oriented segments on spins at point ${\bf r}$ 
\begin{eqnarray}                                                        
S^x(\rvec)&=&S_0 \exp(i\Qvec\cdot\rvec) \cos\Phi(\rvec) \label{eq:sx}\nonumber \\
S^y(\rvec)&=&S_0 \exp(i\Qvec\cdot\rvec) \sin\Phi(\rvec) \label{eq:sy}  
\end{eqnarray}                                                          
where $\Qvec=(\pi,\pi)$ is the AF wave-vector and the phase
$\Phi(\rvec)$ can be
expressed by mapping the Cartesian plane into the complex plane 
$(x,y)\rightarrow z=x+iy$,                     
\begin{equation}\label{eq:phisum}
\Phi(z)={\rm Im} \sum_{i=1}^{N_{seg}}\left\lbrack \ln(z-z_{i,-})-\ln(z-z_{i,+}) 
\right\rbrack .
\end{equation}
Eqs.~(\ref{eq:sy}),(\ref{eq:phisum}) yield 
the angular spin distortion due to the $N_{seg}$ segments 
with $\pm$ ``charges'' at  $z_{i,\pm}$ (${\bf r}_{i,\pm}$). 

By direct integration we obtain that the average spin current in a
square system of dimension $L^2$ in the direction perpendicular to the
``dipole moment'' ${\bf p}_i={\bf r}_+-{\bf r}_-$ is 
$\langle \nabla \Phi(\rvec) \rangle=\pi   {\bf p}_i \times\hat z/L^2
$ with $\hat z$ pointing out of the plane. The macroscopic spin
current is given by $\nabla \Phi(\rvec)_{mac}=\pi  \hat z\times {\bf
  P}$ where 
${\bf P}$ is the macroscopic polarization  in the charge
analogy (${\bf P}={\bf p} N_{seg}/L^2$ for identical dipoles). 
Clearly ${\bf P}$ plays the role of the
ferronematic order parameter. From Eq.~(\ref{eq:sx}) one obtains that 
the macroscopic 
spin current implies an incommensurate
spin response perpendicular to the segments with 
${\bf q}= \nabla \Phi(\rvec)_{mac}$.
This is one of our central results and relates the incommensurability
with the ferronematic order parameter. 

The doping dependence of the order parameter can be estimated by
considering segments (number of sites $=l+1$) which
host $N_c$ charges so that  the filling factor 
$\nu=N_c/(l+1)$. Since $x=N_c N_{seg}/L^2$ is
the concentration of charge carriers one thus finds a linear 
dependence of the incommensurability on doping
\begin{equation}\label{eq:qvec}
q_\perp = 2\pi \frac{x}{2\nu} \frac{l}{l+1} \equiv \pi P.
\end{equation}

It is interesting to remark that if one computes the
incommensurability of infinite stripes consisting of collinear domain
walls (thus without spin current) with filling  $\nu$ \cite{zaa89,lor02} 
it coincides with the incommensurability of finite
ferronematic segments with the same filling and at the same doping, 
i.e. such that the total
length of the segments coincides with the total length of the
stripes. Thus experimentally the two phases are not easily
distinguished in the magnetic channel and the main difference will
arise in the charge channel with  equally spaced stripes producing
Bragg peaks in contrast to diffusive scattering in the case of
segments.

For a set of configurations of macroscopically polarized VA segments 
as shown in Fig. \ref{fig5}(a) 
we now evaluate the  magnetic neutron cross section 
for different dopings which in Fig. \ref{fig6} is 
compared with elastic neutron scattering data 
from Ref.~\onlinecite{waki00}. 
The specific scattering geometry 
(Fig. 2b of Ref.~\onlinecite{waki00}) which is composed of 
two twin domains with population
$2:1$ has been taken into account. This gives rise to the asymmetry of the
spectra since $Q_{AF}$ of the B-twin does not coincide with $Q_{AF}$ of
the A-twin. 
The incommensurate peak position $q_c$ is determined by Eq. \ref{eq:qvec}
and is essentially independent from the segment length
$l$ (i.e. the dipole moment $p$).~\cite{note2} On the other hand the length 
influences  the peak 
width as can be seen in the lowest panel of Fig. (\ref{fig6}).
By decreasing the dipole moment $p$ at fixed doping, the increasing
number of segments decreases the fluctuations of the dipole polarization. 
The strength of the incommensurate response is then favored with respect
to the commensurate one.  
As can be seen from Fig. \ref{fig6} one finds excellent agreement with
the experimental data for segments with $4$ VA pairs 
(i.e. $8$ sites) whereas shorter segments underestimate 
the intensity at $Q_{AF}$ (cf. lower panel of Fig. \ref{fig6}). 

As shown in the inset to Fig.~\ref{fig3} $|E_{bind}|$ decreases
with increasing $|t'/t|$,  
the reduction is, however,  more pronounced for diagonal VA pairs.
This results in a crossing of the alignments as a function
of $N$ for $-0.3 < t'/t < -0.2$. In addition the lattice 
structure can influence the orientation.  This explains the fact that
a diagonal incommensurate spin response has only been observed in
non-superconducting underdoped lanthanum cuprates where $|t'/t|$ 
is small\cite{pav01} and which display a LTO structure in the
underdoped regime. At higher doping an increased fraction of
local octahedral tilts with LTT symmetry \cite{billinge}
may drive a reorientation of the stripe segments from diagonal to the 
copper-oxygen direction at the insulator metal transition $x\approx 0.05$.
In Bi$_2$Sr$_2$CaCu$_2$O$_{8+\delta}$ compounds where $t'/t$ is large 
\cite{pav01} we expect alignment of the segments along the Cu-O direction
resulting in an inequivalence of hole density on the corresponding x- and -y
oxygen sites. It therefore would be interesting to investigate whether this 
feature can account for the  intra-unit cell nematicity observed by
scanning tunneling microscopy \cite{lawler} in these materials.

Apart from the excellent agreement found for the spin structure factor
other facts point to the correctness of our interpretation.
We expect that thermal fluctuations will disorder the dipole
orientation reducing the order parameter until a thermodynamic phase
transition occurs. It is not clear whether the high temperature state will be
only nematic as in Fig.~\ref{fig5}(b) with a second transition at 
higher temperature to an isotropic state  of fully 
disordered dipoles or if the transition will be
directly to the isotropic state. In both cases the incommensurability
should display an order parameter behavior.   
In agreement with this expectation an ordered parameter
like temperature dependence of the  incommensurability has already
been noticed in YBCO.~\cite{hinkov,hau10}

Since the ferronematic state breaks inversion symmetry one 
expects on general grounds\cite{mosto06} that
it will lead to a real ferroelectric distortion, {\it i.e.} to become
multiferroic, through e.g. the inverse Dzyaloshinskii-Moriya
mechanism.~\cite{ser06} Indeed, a ferroelectric 
state has recently been  detected in strongly underdoped
lanthanum cuprates.~\cite{pana} Unfortunately, a small number of free
carriers will make the effect undetectable with capacitance
measurements which may explain why it has  been seen only at very low
temperatures. An appealing possibility would be to look for inversion
symmetry breaking with second harmonic generation (SHG) which does not
require perfect insulating behavior.~\cite{fie04} We expect that the
SHG signal as a function of temperature tracks the incommensurability. 

The macroscopic constant spin current can be considered as an average 
spiral behavior of the spins. In this sense our theory has some
similarity with the proposal of Ref.~\cite{sus05}. However, in our case
the spiral is a collective effect which is slaved by the ferronematic
order of the stripe segments. 

Concluding, we have proposed a new phase for strongly underdoped
cuprates which breaks $C_4$ rotational and inversion symmetry.
Our theory provides a consistent explanation for the elastic incommensurate
response seen by magnetic neutron scattering experiments and reconciles 
it with the lacking signatures of charge order. 
It remains to be seen how the
nematic segments act as seeds which lead to smectic  correlations
(stripes) in some of the cuprate materials.

We thank S. Caprara, R. De Renzi, S. Sanna and P. Carretta  for
insightful discussions and     
S. Wakimoto for sending to us the data shown in Fig.~\ref{fig6}. 
 J. Lorenzana is supported by Italian Institute of Technology-Seed project NEWDFESCM. G. Seibold acknowledges support from the
Deutsche Forschungsgemeinschaft.


\begin{thebibliography}{10}
\bibitem{note2} Note that Eq.~(\ref{eq:qvec}) was derived without
  allowing the background spins to relax. On the other hand the
  explicit calculations  leading to Fig. \ref{fig6} were carried out
  by minimizing the magnetic energy.  
This even more reduces the dependence of $q_c$ on $p$.
\end{thebibliography}
\end{document}